\documentclass{appolb}
\usepackage{epsfig}

\begin{document}
\newcount\eLiNe\eLiNe=\inputlineno\advance\eLiNe by -1
\title{Status of lattice structure function calculations
\thanks{Presented at the DIS2002 International Workshop on Deep Inelastic
        Scattering, Krak\'ow, Poland, April 30 to May 4, 2002.}
}
\author{Stefano Capitani
\address{DESY Zeuthen \\
         John von Neumann-Institut f\"ur Computing (NIC)\\
         Platanenallee 6, 15738 Zeuthen, Germany}
}
\maketitle
\begin{abstract}
Lattice QCD allows computations of moments of structure functions
from first principles. An overview of the present status
of the calculations is given. Recent results and future perspectives
are discussed.
\end{abstract}

\section{Introduction}

The moments of structure functions can be calculated from first principles,
without making any model assumptions, using the non-perturbative techniques 
of lattice QCD. Numerous results have been obtained from lattice computations 
in the last years. They include the calculation of the lowest moments 
of the unpolarized structure functions, of the spin-dependent $g_1$ and $g_2$ 
structure functions and of the $h_1$ transversity structure function. 
Some higher-twist matrix elements have been studied as well.

The novelties of the last couple of years are the first computations 
of structure functions done in full QCD (unquenched), and new proposals 
for the extrapolations to the chiral limit using chiral perturbation theory.
A lot of work however still needs to be done in order to control other 
systematic uncertainties like the continuum limit or the non-perturbative 
renormalization.

\section{Structure functions on the lattice}

It is not possible to compute the structure functions directly on the lattice,
as they describe the physics close to the light cone, and Monte Carlo
simulations are instead done in Euclidean space.
One can however calculate their moments via OPEs which have the general form 
\begin{equation}
\int_0^1 x^n \, {\cal F}_i (x,Q^2) \, \sim \, C_{i,n} (Q^2/\mu^2) 
\cdot \, \langle \, h \, | O_n (\mu)  | \, h \, \rangle .
\label{eq:ope}
\end{equation}
The Wilson coefficients $C_{i,n}$ contain the short-distance physics, 
calculable in continuum perturbation theory. The long-distance physics is 
contained in the hadronic matrix elements, which can be computed doing 
Monte Carlo simulations and then performing a proper lattice renormalization.

The expansions of Eq.~(\ref{eq:ope}) are dominated by operators of twist two.
The moments of the unpolarized quark distributions, of the spin-dependent 
structure function $g_1$, of the $d_2$ component of $g_2$ and of the $h_1$ 
transversity structure function are respectively measured by the 
following operators:
\begin{eqletters}
\begin{eqnarray} 
\langle x^n \rangle &\sim& \langle \, h \, | \,
\overline{\psi} \, \gamma_{\{\mu} D_{\mu_1} \cdots D_{\mu_n\}} \, \psi
\, | \, h \, \rangle , 
\label{eq:unpol_q} \\ 
\langle (\Delta x)^n \rangle &\sim& \langle \, h \, | \, 
\overline{\psi} \, \gamma_5 \gamma_{\{\mu} D_{\mu_1} \cdots D_{\mu_n\}} \, \psi
\, | \, h \, \rangle , \\
\langle x^n \rangle_{d_2} &\sim& \langle \, h \, | \, 
\overline{\psi} \, \gamma_5 \gamma_{[\mu} D_{\{\mu_1} D_{\mu_2]}
\cdots D_{\mu_n\}} \, \psi
\, | \, h \, \rangle , \\
\langle (\delta x)^n \rangle &\sim& \langle \, h \, | \,
\overline{\psi} \, \gamma_5 \sigma_{\mu\{\mu_1} D_{\mu_2} 
\cdots D_{\mu_n\}} \, \psi
\, | \, h \, \rangle .
\end{eqnarray}
\end{eqletters}
The moments of the unpolarized gluon distribution,
\begin{equation}
\langle x^n \rangle_g \sim \langle \, h \, | \,
\sum_\rho \Tr \, (F_\mu^{\ \rho} D_{\mu_2} \cdots D_{\mu_n} F_{\rho\mu_1})
\, | \, h \, \rangle ,
\end{equation}
which in the flavor singlet case mix with the moments in 
Eq.~(\ref{eq:unpol_q}), have been quite hard to compute on the lattice 
up to now, and for this reason only non-singlet quark distributions 
are generally considered.

Structure functions calculations on the lattice have been done with 
Wilson fermions, which however break chiral symmetry. This causes 
an additive mass renormalization even for vanishing bare masses, 
and a heavy lattice pion ($m_\pi \sim$ 500 MeV). 
Therefore one needs to perform extrapolations to the chiral limit, 
and this systematic error has to be controlled.~\footnote{Recent 
formulations of chiral fermions on the lattice, called Ginsparg-Wilson 
(in particular: overlap, domain-wall and fixed-point fermions), possess 
exact chiral symmetry also at finite lattice spacing and thus do not suffer
from this problem.}

Extrapolations to the continuum limit $a \rightarrow 0$ need also to be done.
To get a faster convergence to the continuum and decrease 
the systematic error arising from the finiteness of the lattice spacing,
$O(a)$ improvement has been implemented in most cases: 
the contributions of $O(a)$ are removed so that one has
\begin{equation}
\left< p \left| \widehat{\cal O}_{L} \right| p' \right>_{MC}=
a^{d} \left[ \left< p \left| \widehat{\cal O} \right| p' 
\right>_{phys} + O(a^2) \right] .
\end{equation} 
This is achieved in on-shell matrix elements by adding a counterterm 
to the Wilson action, $\Delta S^f_I = c_{sw} \cdot {\rm i} g_0 a^{4} \,
r/4a \, \sum_{x,\mu \nu} \overline{\psi} (x)
\sigma_{\mu \nu} F_{\mu \nu}^{\rm clover}(x) \psi (x)$. 
The improvement coefficient $c_{sw}$, which has to be exactly tuned so as 
to cancel all $O(a)$ contributions coming from the action, is now well known.
The various operators have to be improved as well, and this is achieved 
by adding as counterterms bases of higher-dimensional operators with the same 
symmetries as the original unimproved ones: 
$O^{\rm imp} = (1+b_O \, a m) \, O + a \sum_i c_i \, \widetilde{O}_i$ (with
$\dim (\widetilde{O}_i) = \dim (O) + 1$), and determining the coefficients
$b_O$ and $c_i$.
 
After the matrix elements have been simulated with Monte Carlo algorithms, 
they need to be renormalized from the lattice to the continuum:
\begin{equation}
\langle O_i^{\rm cont} \rangle = \sum_j \Bigg( \delta_{ij} 
-\frac{g_0^2}{16 \pi^2} \Big( R_{ij}^{\rm lat} -R_{ij}^{\rm cont} \Big) \Bigg) 
\cdot \langle O_j^{\rm lat} \rangle .
\end{equation}
Since Lorentz and (for Wilson fermions) chiral symmetry are broken on the 
lattice, mixing of operators under renormalization is more complicated than 
in the continuum. Much effort has been spent on computing perturbative 
renormalization factors, which are now known for the lowest three moments 
of all structure functions in the case of Wilson (often in the $O(a)$ improved
theory) and of overlap fermions. For renormalization calculations see 
Ref.~\cite{pertrenorm} and references therein. 
The relevant perturbative renormalization factors have been calculated 
also for some classes of higher-twist operators~\cite{ht}.

Non-perturbative renormalization has instead been quite hard to do in the 
standard Wilson case, where it is difficult to find a plateau for extracting 
the signal. This is not the case in the Schr\"odinger Functional approach,
where one has a better control over many systematic errors, 
although the computation of the running of renormalization factors 
is quite tedious~\cite{SchFun}.

For reasons of computing power, structure functions have been studied 
mostly in the quenched approximation, and only recently full QCD computations 
have become feasible. On the lattice the (Grassmann) fermion variables are 
integrated out, and in full QCD one uses the partition function
\begin{equation}
Z = \int {\cal D} U \, \, \det (\not{\hspace{-0.12cm}D}[U]+m_q) 
\, \, {\rm e}^{-S_g[U]} \quad \quad (U_\mu = {\rm e}^{{\rm i}a g_0 A_\mu}) .
\end{equation}
Quenching instead amounts to putting 
$\det (\not{\hspace{-0.20cm}D}[U]+m_q) = 1$ in the simulations. 
In physical terms, this means that there are no sea quarks in the 
calculations: the internal quark loops are neglected. Although it looks
quite drastic, in many cases this does not turn out to be a bad approximation.

\section{Results}

Various collaborations have been studying structure functions on the lattice 
in the last years: the QCDSF Collaboration, the LHPC Collaboration, and a 
collaboration which uses the Schr\"odinger Functional scheme.
See Refs.~\cite{QCDSF,LHPC,SchFun} respectively for recent works and
further references.

The major advancement in the last couple of years has been the computation 
of the structure functions in full QCD. 
The first studies with dynamical fermions have shown that for various 
structure functions of the $u$ and $d$ quarks there are no statistically 
significant differences between quenched and full QCD results. This was 
discovered by the LHPC Collaboration and then confirmed by the QCDSF 
Collaboration. We then have two calculations in full QCD with the same 
message: the results for quenched and unquenched moments are nearly equal, 
and the observed discrepancies of lattice results with experiment 
are still there. Quenching had been previously conjectured as the cause 
of these discrepancies, but evidently this is not the case, at least 
for the values of quark masses attainable at present, which are 
unfortunately quite large.

Usually only the lowest three or so moments of the various structure functions
can be calculated, due to increasing computational difficulties and mixing 
problems. The largest discrepancies concern unpolarized quark distributions.
The spin-dependent $g_1$ and $g_2$ structure functions have been extensively 
studied as well, and also the axial charge $g_A = \Delta u - \Delta d$, 
whose result is not so far from experiment. 
The $g_2$ structure function with Wilson fermions presents a mixing problem 
due to chirality breaking:~\footnote{With Ginsparg-Wilson fermions this mixing
is thus forbidden.} there is a mixing with lower-dimensional operators, 
which gives rise to power divergences $\sim 1/a^n$ in the continuum limit.
The $h_1$ transversity structure function, whose lowest moment
is the tensor charge, has also been studied on the lattice. 
A recent lattice result for this quantity, which has never been measured, is 
$\delta u - \delta d = 1.21(4)$~\cite{transv}. An experimental measure of the 
$h_1$ transversity structure functions would therefore be quite interesting.

The results coming from lattice QCD are getting more and more precise,
so it would also be useful to have more precise measurements of parton 
distributions and above all a careful analysis of their errors. 
Some first studies in this direction have been presented at this 
conference~\cite{param}.

Nevertheless, on current lattices there are still some practical limitations.
One of them is given by the difficulty to compute disconnected diagrams 
(\ie connected, but only by gluon lines). Only exploratory studies 
have been made so far, and these diagrams have not yet been included in
lattice computations. They are however flavor independent, 
and therefore they do not contribute to the differences between $u$ and $d$ 
structure functions (for $m_u=m_d$, which is the case).
This means that quantities like $g_A = \Delta u -\Delta d$
and $\langle x^n \rangle_{u-d}$ do not receive contributions
from disconnected diagrams.

Another limitation is given by the extrapolations to the chiral limit.
Chiral and continuum extrapolations are performed using the fit formula
\begin{equation}
A + B \, m_\pi^2 + C \, a^2 ,
\label{eq:fit}
\end{equation}
and doing just the naive chiral extrapolations in $m_\pi^2$ could also turn 
out to be an explanation for the discrepancies. The extrapolations to the 
continuum limit are quadratic in the lattice spacing when improved 
fermions are used.

There have also been quenched calculations which use the Schr\"odinger 
Functional. This is a finite volume scheme (with Dirichlet boundary conditions
on the time direction) in which it is possible to do simulations at very small 
quark masses, and the pion is much lighter than in the standard Wilson case.
The renormalization scale is identified with the inverse size of the lattice, 
$\mu = 1/L$, which allows one to apply finite-size scaling techniques;
there is no need to go to the infinite volume limit.
In the Schr\"odinger Functional calculations the renormalization 
of operators can be done non-perturbatively, at scales extending over 
some orders of magnitude.  
The results for the lowest moment of the unpolarized structure function 
of the pion, in the $\overline{\rm{MS}}$ scheme at the renormalization scale 
$\mu=$ 2.4 GeV, is $\langle x \rangle = 0.30 \pm 0.03$,
while the experimental number is $\langle x \rangle = 0.23 \pm 0.02$.

\section{Chiral extrapolations}

It has recently been suggested that extrapolations of the lattice data 
using chiral perturbation theory could solve the discrepancy with experiment.

It seems that the pion cloud of the nucleon is not adequately described 
by current lattices. The IR behavior of pion loops generates
chiral logs $\sim m_\pi^2 \, \ln m_\pi^2$, and using chiral perturbation 
theory~\cite{chpt} one gets for the unpolarized case 
$\langle x^n \rangle_{u-d} \sim A_n \, \Big[ 1 
- (3g_A^2+1)/(4\pi f_\pi )^2 \cdot m_\pi^2 \, \ln m_\pi^2 \Big]$. 
Introducing in this non-analytic term a phenomenological cutoff $\Lambda$
(the size of the source generating the pion cloud) 
one then arrives at the fit formula~\cite{chirextr}
\begin{equation}
\langle x^n \rangle_{u-d} = A_n \, \Big[ 1 
- \frac{(3g_A^2+1)}{(4\pi f_\pi )^2} \, m_\pi^2 \, 
\ln \Big(\frac{m_\pi^2}{m_\pi^2 + \Lambda^2} \Big) \Big] 
+ B_n \, m_\pi^2 + C_n \, a^2.
\label{eq:chptfit}
\end{equation}

It looks like phenomenological extrapolations which use chiral perturbation 
theory could solve the discrepancy with experiment. However, they need to be 
better investigated, and although they have the potential to explain the 
discrepancy, it is still too soon to say something certain.~\footnote{A 
recent calculation~\cite{polchirextr} seems also to show that in the polarized
case the next order corrections give a fit formula which deviates very little 
from the naive extrapolations.}

So far we are able to reproduce the phenomenological results only 
at posteriori. The corresponding values of $\Lambda$ lie between 
300 MeV and 700 MeV, so $\Lambda$ does not seem at present 
to have some predictive power.
Moreover, the currently available lattice data do not even 
discriminate between naive chiral fits, Eq.~(\ref{eq:fit}),
and chiral perturbation theory fits, Eq.~(\ref{eq:chptfit}).
The problem is that in present simulations the pions are not sufficiently 
light. A smaller pion mass is needed ($m_\pi <$ 250 MeV) in order that 
the parameters of the chiral expansions can be well determined on the lattice.
For such determinations a computing power of about 8 Teraflops for one year 
is required~\cite{chirextr}. The next generation of computers, which will 
come in a couple of years, should then be able to perform these calculations.

For the pion cloud of the proton to be adequately included in the lattice box 
and properly measured in Monte Carlo simulations, the pion correlation length 
should also be smaller than the lattice size, otherwise it will not be fully 
contained in the lattice.
The length of a spatial dimension for currently avaliable lattices is
about 2 fermi, so it would be very useful to do simulations also with larger 
physical volumes and minimize finite volume effects.~\footnote{We should 
also mention that the lattice results for elastic form factors seem to have 
discrepancy problems similar to the structure functions, particularly in the 
case of the magnetic form factor. It is likely that in this case too the pion 
cloud is not being properly simulated and finite-size effects are present.}

The pion cloud will be the focus of many future lattice investigations.

\section{Higher twist}

There are a few lattice results regarding higher-twist corrections for 
the pion and the nucleon~\cite{ht}. They show that, for the twist-four 
contributions coming from 4-fermion operators like 
$\sum_A \, \, \overline{\psi} \gamma_\mu \gamma_5 \, t^A \psi \cdot 
\overline{\psi} \gamma_\nu \gamma_5 \, t^A \psi$,
the $1/Q^2$ power corrections are, 
at least in these particular cases, quite small.

The calculations give, for a particular contribution ($I=2$) to the twist-4 
matrix element of the first moment of the unpolarized pion distribution,
\begin{equation}
\int_0^1 dx \, F_2 (x,Q^2) \Big|_{I=2} = 1.67(64) \cdot \alpha_S (Q^2) \cdot
\frac{f_\pi^2}{Q^2}, 
\end{equation}
and for the proton, for a particular class of operators 
(flavor ${\bf 27}$, $I=1$),
\begin{equation}
\int_0^1 dx \, F_2 (x,Q^2) \Big|_{{\bf 27},I=1} = -0.0006(5) 
\cdot \alpha_S (Q^2) \cdot \frac{m_P^2}{Q^2}.
\end{equation}
These special flavor and isospin combinations are chosen to avoid mixing 
with lower-dimensional operators. The above twist-4 numbers are much 
smaller than the twist-2 lattice matrix element (at $Q^2=$ 4 GeV$^2$) of the 
pion, $\int_0^1 dx \, F_2 (x,Q^2) = 0.152(7)$,
and of the proton, which is about $0.14$.

We point out that these studies, contrary to leading-twist operators, 
are not complete and systematic. Mixing problems have so far limited 
the calculations only to a restricted class of operators.

\section{Perspectives}

The lattice calculations of structure functions are getting more and more 
refined. However, the physics of the pion cloud must be properly included, 
and simulations on lattices of large physical volumes as well as proper 
extrapolations to the chiral limit will need to be done.

Ginsparg-Wilson fermions possess exact chiral symmetry and could help 
to study the chiral limit. Calculations with the Schr\"odinger Functional 
will also be useful for a better understanding of the approach
to the chiral limit.

Contributions of higher-twist operators to moments of parton distributions 
seem to be quite small, but they have not been systematically studied.
They will continue to be a challenge for lattice QCD for a few more years.

Disconnected diagrams will have to be included in the simulations.

An almost virgin territory for the lattice is the study of the momentum 
and spin distributions of the gluon, which so far have been plagued 
in Monte Carlo simulations by large statistical errors.
It would be quite interesting to know them.
The lattice distributions of sea quarks are also unknown.

\end{document}